\documentclass[12pt]{article}

\usepackage{graphicx}

\usepackage{pslatex}
\usepackage{epstopdf}
\def\la{{\langle}}
\def\ra{{\rangle}}
\newcommand{\beq}{\begin{equation}}
\newcommand{\eeq}{\end{equation}}
\newcommand{\beqa}{\begin{eqnarray}}
\newcommand{\eeqa}{\end{eqnarray}}

\begin{document}
\baselineskip 14pt

\begin{center}
\vspace*{2cm}
{\Large\bf Discrimination of measurement contexts in quantum mechanics\vspace*{1cm}\\}
%
%

{\large R. Sala Mayato}$^{a,b}$\footnote{E-mail address: 
rsala@ull.es } and {\large J. G. Muga}$^{a,c}$\\

$^a${\it Max Planck Institute for the Physics of Complex Systems,\\
N\"othnitzer Str. 38 D-01187 Dresden, Germany}\\

$^b${\it Departamento de F\'\i sica Fundamental II and IUdEA,\\
Universidad de La Laguna, La Laguna, 38204, S/C de Tenerife, Spain}\\

$^c${\it Departamento de Qu\'\i mica F\'\i sica,\\
UPV/EHU, Apartado 644, Bilbao 48080, Spain}

\date{\today}


\end{center}

\pagestyle{plain}
\begin{center}
{\bf Abstract}
\end{center}
\begin{small}
We demonstrate    
that it is possible to discern the way that has been followed to measure a quantum observable that can be expressed in terms of different products of observables, whereas no such discrimination is possible by assigning predetermined values. Specifically 
we show how to distinguish different  routes (contexts) to measure $C=AB=A'B'$, when $C,A,B$ and $C,A',B'$ commute with each other, but $A$ and $B$ do not commute with $A'$ and $B'$. 

PACS: 03.65.Ca, 03.65.Ta
\end{small}
\vspace*{1cm}

%
%

\section{Introduction}



The point of the Bell-Kochen-Specker (Bell-KS) theorem \cite{Bell,KS} is that 
we have to give up the simple, classical assumption that measuring a quantity tells us the value the quantity had {\it before} the measurement. The results of the 
measurement are ``contextual'', in the sense that they depend on the 
experimental arrangement. 
A particularly simple proof of the Bell-KS theorem
is due to Peres and Mermin \cite{Peres,Mermin}, based on a set of nine observables (Mermin's square).   
Quantum contextuality experiments with sequential measurements based on the Peres-Mermin proof have been proposed by Cabello \cite{Cabello}.

Our purpose here is to bring forward   
an aspect of contextuality that has been overlooked, namely, 
that it is quantally possible to discern the way that has been followed to measure an observable that can be expressed in terms of different products of observables, whereas no such discrimination is possible by assigning predetermined values. From the experimental point of view, ``measure an observable'' in this paper is to be understood as repeating the experiment many times so that the expectation value and the statistical distribution of eigenvalues may be obtained. One single realization of the experiment with just one eigenvalue as outcome will be referred to as an ``individual measurement''. 

In order to understand our main result, it is important to distinguish between what traditionally appears in textbooks as {\em von Neumann measurement} \cite{vN} from {\em L\"uders measurement} \cite{Lu}.    
The projection postulate as commonly used nowdays is in fact due to G. L\"uders \cite{Lu}. For observables with degenerate eigenvalues his formulation differs from that of von Neumann \cite{vN}. In the degenerate case these two formulations represent different measurement processes for which the final states are, in general, also different, as shown next in detail.
For a clarifying discussion see \cite{Lu}. 

The projection postulate intends to describe the effects of an ideal measurement on the state of a system, and it has been widely regarded as a useful tool. 
Let $O$ be an observable, and suppose it has discrete eigenvalues $o_1,o_2,o_3,...$ with associated degeneracies $g_1,g_2,g_3,...$ respectively. 
Asume that the system is prepared in the state described by the density matrix $\rho$. Then, the measuring operator $O$ can yield any of the results $o_1,o_2,o_3,...$ and the measurement process results in a new density matrix, $\rho'$. 
In L\"uders' measurement,
\beq\label{Lum}
\rho'=\sum_{n} P_n\rho P_n\, ,
\eeq
where $P_n$ is the projection operator into the subspace associated with the eigenvalue $o_n$,
\beq
P_n=\sum_{i=1}^{g_n}|o_n^i\ra\la o_n^i|\, ,
\eeq
where the vectors $|o_n^i\ra$ for fixed $n$ are the degenerate eigenvectors associated to the eigenvalue $o_n$.

In von Neumann's measurement,
\beq\label{vNm}
\rho'=\sum_{n,i} P_n^i\rho P_n^i\, ,
\eeq
where $P_n^i$ is the projection operator associated with each eigenvector of the operator $O$, i.e.,
\beq
P_n^i=|o_n^i\ra\la o_n^i|\, .
\eeq
When the eigenvalues of the operator we are measuring are not degenerate, both measurement processes reach the same final state. 
\section{Equivalence and inequivalence of operators}
In Mermin's square \cite{Mermin},
\vspace*{0.5cm}
\beq
\begin{array}{lll}
\sigma_{1x} & \sigma_{2x} & \sigma_{1x}\sigma_{2x} \\
\sigma_{2y} & \sigma_{1y} & \sigma_{1y}\sigma_{2y} \\
\sigma_{1x}\sigma_{2y} & \sigma_{2x}\sigma_{1y} & \sigma_{1z}\sigma_{2z}\, ,
\end{array}
\vspace*{0.5cm}
\eeq
%
%
the observables are the Pauli matrices, with eigenvalues $\pm 1$, for two independent spin$-1/2$ particles. The nine observables are arranged in groups of three columns and three rows, and within each of them they are mutually commuting. The product of the three observables in the column on the right is $-1$, and the product of the three observables in the other two columns and all three rows is $+1$. It is easy to check that it is impossible to associate with each observable preexisting values, $+1$ or $-1$, in such a way that they verify the identities satisfied by the observables themselves.


In this letter, using only one of the columns and one of the rows in Mermin's square, we pose a question for which the association of preexisting values to some magnitudes gives no answer while standard quantum mechanics provides one. 
Let us consider the operator $C=AB=A'B'$ such that $A$, $B$ and $C$ commute with each other, as do $A'$, $B'$ and $C$, but $A$ and $B$ do not commute with $A'$ and $B'$. We shall use for concreteness the first column and the third row in Mermin's square, i.e.,
\vspace*{0.5cm}
\beq
\begin{array}{lcc}
A=\sigma_{1x} &  & \\
B=\sigma_{2y} &  &  \\
C=\sigma_{1x}\sigma_{2y} & B'=\sigma_{2x}\sigma_{1y} & A'=\sigma_{1z}\sigma_{2z}\, .
\end{array}
\vspace*{0.5cm}
\eeq
With this choice, indeed, 
$C=AB=A'B'$. Also, in agreement with the premises, $A$, $B$ and $C$ commute with each other, as  $A'$, $B'$ 
and $C$ do, but $A$ and $B$ do not commute with $A'$ and $B'$\footnote{We could  find the same results using any other pair of rows and columns except the last column.}. 

In principle, we could measure  
$C$ by three different routes, i.e. within three different contexts, namely: by measuring $C$ directly, by measuring $AB$ (say first $B$ and then $A$, or viceversa, since they commute), or by measuring $A'B'$ (again, one after the other one).
The question announced above is this: would it be possible to distinguish after the measurement the route (the context) followed to measure $C$? 

Suppose we assign preexisting values to each of the magnitudes that verify $C=AB=A'B'$, such that they take values $+1$ and $-1$. The possible cases are given in the following table,
\vspace*{0.5cm}
\beq
\begin{array}{ccccc}
A & B & A' & B' & C \\
+1 & +1 & +1 & +1 & +1 \\
+1 & +1 & -1 & -1 & +1 \\
-1 & -1 & +1 & +1 & +1 \\
-1 & -1 & -1 & -1 & +1 \\
-1 & +1 & -1 & +1 & -1 \\
+1 & -1 & +1 & -1 & -1 \\
+1 & -1 & -1 & +1 & -1 \\
-1 & +1 & +1 & -1 & -1\, .
\end{array}
\eeq
%
If the measurements revealed the values of this table, it would be impossible to assert if the magnitude $C$ was measured directly, via $AB$ or via $A'B'$.

%
%

Now, let us address the question from the point of view of standard quantum mechanics.
Assume an initial state given by 
\beq\label{initial}
|\psi\ra=\alpha|++\ra +\beta|+-\ra +\gamma|-+\ra +\delta|--\ra\, ,
\eeq
where $\{|++\ra,|+-\ra,|-+\ra,|--\ra\}$ are the eigenvectors of $\sigma_{1z}$ and $\sigma_{2z}$. The corresponding density operator  would be $\rho_i = |\psi\ra\la\psi|$, where its coefficients are given in Appendix 1.

The measurements through the three different arrangements (contexts) give the same statistical results for $C$, but  
they also produce three different final states: 
\beqa\label{Crho}
\rho_{C}&=&\frac{1}{4}[c_{11}|\phi_1\ra\la\phi_1| + c_{22}|\phi_2\ra\la\phi_2| + c_{33}|\phi_3\ra\la\phi_3| + c_{44}|\phi_4\ra\la\phi_4|\nonumber\\
&+& c_{14}|\phi_1\ra\la\phi_4| + c_{41}|\phi_4\ra\la\phi_1| + c_{23}|\phi_2\ra\la\phi_3| + c_{32}|\phi_3\ra\la\phi_2|]
\eeqa
measuring $C$ directly,
\beq\label{ABrho}
\rho_{AB}=\frac{1}{4}[c_{11}|\phi_1\ra\la\phi_1| + c_{22}|\phi_2\ra\la\phi_2| + c_{33}|\phi_3\ra\la\phi_3| + c_{44}|\phi_4\ra\la\phi_4|]
\eeq
measuring $AB$,  and
\beqa\label{primarho}
\rho_{A'B'}&=&\frac{1}{4}[d_{11}|\phi_1\ra\la\phi_1| + d_{11}|\phi_2\ra\la\phi_2| + d_{22}|\phi_3\ra\la\phi_3| + d_{22}|\phi_4\ra\la\phi_4|\nonumber\\
&+& d_{14}|\phi_1\ra\la\phi_4| + d_{14}|\phi_4\ra\la\phi_1| + d_{23}|\phi_2\ra\la\phi_3| + d_{23}|\phi_3\ra\la\phi_2|]
\eeqa
measuring $A'B'$,
where we have used the set 
of common eigenvectors of $\{A,B,C\}$, 
$\{|\phi_1\ra,|\phi_2\ra,|\phi_3\ra,|\phi_4\ra\}$. The relations among the different bases and the explicit expressions of coeficients  $c$'s and $d$'s, are given in Appendix 1.

The difference between the density operators (\ref{Crho}) and (\ref{ABrho}) that describe the state after the 
measurement, is clear: in $\rho_{AB}$ the cross terms (the coherences) are absent.  
In both cases, we have applied L\"uders recipe (\ref{Lum}). However, 
the result of applying L\"uders rule twice (associated with measuring first $A$ and then $B$) 
is equivalent to applying von Neumann's recipe (\ref{vNm}) directly on $C$.  von  Neumann's measurement removes the cross terms among the different eigenvectors of the basis.

%
%


To distinguish the routes we may use the different final states: the expectation values of the operators will be in general different too. In particular, 
\beqa
\la A\ra_C &=&\la A\ra_{AB}=\Re(r_{13})+\Re(r_{24})\nonumber\\
\la A \ra_{A'B'}&=&0\label{mva}\\
%
\la B\ra_C &=&\la B\ra_{AB}=-\Im(r_{12})-\Im(r_{34})\nonumber\\
\la B\ra_{A'B'}&=&0\label{mvb}\\
%
\la A'\ra_C &=&\la A'\ra_{A'B'}=-\Im(r_{14})-\Im(r_{23})\nonumber\\
\la A'\ra_{AB}&=&0\label{mva'}\\
%
\la B'\ra_C &=&\la B'\ra_{A'B'}=r_{11}-r_{22}-r_{33}+r_{44}\nonumber\\
\la B'\ra_{AB}&=&0\label{mvb'}\\
%
\la C\ra_C &=&\la C\ra_{AB}=\la C\ra_{A'B'}=-\Im(r_{14})+\Im(r_{23})\label{mvc}\, ,
\eeqa
where the subscripts $C$, $AB$ or $A'B'$ mean that the expectation value is calculated with 
$\rho_{C}$, $\rho_{AB}$ or  $\rho_{A'B'}$ respectively.
It is remarkable that the mean values of $A$ and $B$ are strictly zero via $A'B'$ and, 
conversely, the mean values of $A'$ and $B'$ are zero via $AB$. 
We shall assume in the following that $\la B\ra\ne0$ and $\la B'\ra\ne0$ for the initial state. If they were zero we would prepare a different state.   

Now, suppose we measure $C$ performing 1000 individual measurements in identical conditions, and we want to know if such measurement was made via $AB$, $A'B'$ or $C$. Clearly we cannot discern this from the mean value of $C$, because it is route-independent (see (\ref{mvc})). However, if we take 500 of our systems and evaluate the mean value of observable $B$ (the same argument can be made with $A$), by measuring such observable, and get zero then, the measurement of $C$ has been done necessarily via $A'B'$ (see (\ref{mvb})). $\la B\ra_C =\la B\ra_{AB}=-\Im(r_{12})-\Im(r_{34})$ cannot be equal to zero because of our initial condition, $\la B\ra\ne0$.
On the other hand, if such expectation value is not zero, the mesurement had to be done via $AB$ or $C$ (see (\ref{mvb})). To discriminate in this case we just have to find the mean value of $B'$ on the other 500  systems: if it is different from zero,  then the measurement was made via $C$ and, if it is zero, it was made via $AB$, see (\ref{mvb'}).


As a simple example consider the initial state
\beq
|\psi\ra=\frac{1}{\sqrt3}(|++\ra -i|+-\ra +|--\ra)\, .\nonumber\\
\eeq
Following the general approach explained above, we have three different final states
measuring $C$ directly, $AB$ or $A'B'$:   
\beqa
\rho_{C}&=&\frac{1}{12}(|\phi_1\ra\la\phi_1| + 5|\phi_2\ra\la\phi_2| + |\phi_3\ra\la\phi_3| + 5|\phi_4\ra\la\phi_4|\nonumber\\
&+& (1+2i)|\phi_1\ra\la\phi_4| + (1-2i)|\phi_4\ra\la\phi_1| + |\phi_2\ra\la\phi_3| +|\phi_3\ra\la\phi_2|)\nonumber\\
%
%
\rho_{AB}&=&\frac{1}{12}(|\phi_1\ra\la\phi_1| + 5|\phi_2\ra\la\phi_2| + |\phi_3\ra\la\phi_3| + 5|\phi_4\ra\la\phi_4|)\nonumber\\
%
%
\rho_{A'B'}&=&\frac{1}{4}+\frac{1}{12}(|\phi_1\ra\la\phi_4| + |\phi_4\ra\la\phi_1| + |\phi_2\ra\la\phi_3| +|\phi_3\ra\la\phi_2|).\nonumber
\eeqa
%
The mean values 
from Eq. (\ref{mva}) to (\ref{mvc}) are 
\beqa
\la A\ra_C &=&\la A\ra_{AB}=\la A \ra_{A'B'}=0\nonumber\\
\la B\ra_C &=&\la B\ra_{AB}=-\frac{1}{3}\,,\;\;\la B\ra_{A'B'}=0\nonumber\\
\la A'\ra_C &=&\la A'\ra_{A'B'}=\la A'\ra_{AB}=0\nonumber\\
\la B'\ra_C &=&\la B'\ra_{A'B'}=\frac{1}{3}\,,\;\;\la B'\ra_{AB}=0\nonumber\\
\la C\ra_C &=&\la C\ra_{AB}=\la C\ra_{A'B'}=0\, .\nonumber
\eeqa
With the mean values of $B$ and $B'$ we can discriminate among the different routes
followed. 


Remark 1: The decomposition $C=AB=A'B'$ verifying that $A$, $B$ and $C$ commute with each other, as do $A'$, $B'$ and $C$, but such that $A$ and $B$ do not commute with $A'$ and $B'$, is only possible if $C$ has a degenerate spectrum. 

Remark 2: To demonstrate contextuality the equality $C=AB$ is enough because, even if the statistical results for the measurement of both sides of this expression are equal, the final state they induce is not necessarily the same.  This provides different expectation values for some observables. Similar results are found using addition instead of products of observables, as shown in Appendix 2.

Remark 3: If predetermined values existed, as in the table above, 
no distinction of route could be performed, even if we did extra measurements to calculate other expectation values. 
\section{Conclusions}
We have shown a striking and simple example of quantum contextuality. 
The route followed to measure an observable expressed by different  operator products, corresponding to different experimental arrangements,  leaves a footprint that can be identified after the measurement. No such tracing back is possible if the measurement merely reveals preexisting values.   

Even without recourse to preexisting values, the result is somehow paradoxical.  
If $C$ is {\it equal} to $AB$, how can we distinguish between $C$ and $AB$? The crux of the matter is that ``equal'' here, as in most equations in physics, does not mean ``identical to'' in all possible respects. Otherwise equations would become useless and tautological.     
In the operator equality $C=AB$, the equal sign means operationally that measurements of $C$ and $AB$ provide the same statistics for the eigenvalues $c$ and products $ab$. $A$, $B$, and $C$ are different objects
implying quite different operations and resulting states, even if they commute! Commutativity adds to the puzzle because, since   
the ordering does not matter, we can measure 
first $A$ and then $B$ or viceversa, with the same final state. Thus the equality $AB=BA$ is of a stronger type
than the equality in $C=AB$. This polysemic character of the equality sign  feeds the paradox, which dissolves away by strictly applying the standard rules 
and recognizing that commuting operators may in summary be equivalent with respect to the statistical results of the measurements but not necessarily with respect to the final state that they induce.       

Finally, we can set interesting goals related to this work: 
one is the possibility to  
implement these measurements in real experiments \cite{Kirchmair,Bartosik,Amselem};  
a second one is to design a test to check whether an apparatus performs a L\"uders or 
a von Neumann measurement. 


\section{Acknowledgemnts}
We thank A. Ruschhaupt,  M. Appleby and G. Hegerfeldt for useful discussions.
We aknowledge funding by Ministerio de 
Ciencia e Innovaci\'on (Grants No. FIS2009-12773-C02-01
and FIS2010-19998) and 
the Basque Government (Grant IT472-10).

\section {Appendix 1}
\begin{itemize}
\item The density operator associated with the state given in (\ref{initial}) is 
\beqa\label{initialrho}
\rho_i &=& |\psi\ra\la\psi|\nonumber\\ 
&=& r_{11}|++\ra\la++| + r_{12}|++\ra\la+-| + r_{13}|++\ra\la-+| + r_{14}|++\ra\la--|\nonumber\\
&+& r_{12}^*|+-\ra\la++| + r_{22}|+-\ra\la+-| + r_{23}|+-\ra\la-+| + r_{24}|+-\ra\la--|\nonumber\\
&+& r_{13}^*|-+\ra\la++| + r_{23}^*|-+\ra\la+-| + r_{33}|-+\ra\la-+| + r_{34}|-+\ra\la--|\nonumber\\
&+& r_{14}^*|--\ra\la++| + r_{24}^*|--\ra\la+-| + r_{34}^*|--\ra\la-+| + r_{44}|--\ra\la--|\, ,
\eeqa
and the coefficients in (\ref{initial}) are related to the coefficients in (\ref{initialrho}): $r_{11}=|\alpha|^2$, $r_{22}=|\beta|^2$, $r_{33}=|\gamma|^2$, $r_{44}=|\delta|^2$, $r_{12}=\alpha\beta^*$, $r_{13}=\alpha\gamma^*$, $r_{14}=\alpha\delta^*$, $r_{23}=\beta\gamma^*$, $r_{24}=\beta\delta^*$, $r_{34}=\gamma\delta^*$.

\item The common eigenvectors of $\{A,B,C\}$ are 
\beqa
|\phi_1\ra &=&\frac{1}{2}{|++\ra+i|+-\ra+|-+\ra+i|--\ra}\nonumber\\
|\phi_2\ra &=&\frac{1}{2}{|++\ra-i|+-\ra+|-+\ra-i|--\ra}\nonumber\\
|\phi_3\ra &=&\frac{1}{2}{|++\ra+i|+-\ra-|-+\ra-i|--\ra}\nonumber\\
|\phi_4\ra &=&\frac{1}{2}{|++\ra-i|+-\ra-|-+\ra+i|--\ra}\, ,
\eeqa
and their associated eigenvalues $a$, $b$ and $c$ are given by
\beq
\begin{array}{cccc}
 & a & b & c\\
|\phi_1\ra & +1 & +1 & +1\\
|\phi_2\ra & -1 & -1 & +1\\
|\phi_3\ra & -1 & +1 & -1\\
|\phi_4\ra & +1 & -1 & -1\, .
\end{array}
\eeq

\item The common eigenvectors of $\{A',B',C\}$ take the form 
\beqa
|\psi_1\ra &=&\frac{1}{2}{|++\ra+|+-\ra+|-+\ra+|--\ra}\nonumber\\
|\psi_2\ra &=&\frac{1}{2}{|++\ra-|+-\ra+|-+\ra-|--\ra}\nonumber\\
|\psi_3\ra &=&\frac{1}{2}{|++\ra+|+-\ra-|-+\ra-|--\ra}\nonumber\\
|\psi_4\ra &=&\frac{1}{2}{|++\ra-|+-\ra-|-+\ra+|--\ra}\, ,
\eeqa
and the associated eigenvalues $a'$, $b'$ and $c$ are given by
\beq
\begin{array}{cccc}
 & a' & b' & c\\
|\psi_1\ra & +1 & +1 & +1\\
|\psi_2\ra & -1 & -1 & +1\\
|\psi_3\ra & -1 & +1 & -1\\
|\psi_4\ra & +1 & -1 & -1\, .
\end{array}
\eeq

\item The c's and d's in expresions (\ref{Crho}), (\ref{ABrho}), and  (\ref{primarho}) are 
\beqa
c_{11}&=&1-2\Im(r_{12})+2\Re(r_{13})-2\Im(r_{14})+2\Im(r_{23})+2\Re(r_{24})-2\Im(r_{34})\nonumber\\
c_{22}&=&1+2\Im(r_{12})+2\Re(r_{13})+2\Im(r_{14})-2\Im(r_{23})+2\Re(r_{24})+2\Im(r_{34})\nonumber\\
c_{33}&=&1-2\Im(r_{12})-2\Re(r_{13})+2\Im(r_{14})-2\Im(r_{23})-2\Re(r_{24})-2\Im(r_{34})\nonumber\\
c_{44}&=&1+2\Im(r_{12})-2\Re(r_{13})-2\Im(r_{14})+2\Im(r_{23})-2\Re(r_{24})+2\Im(r_{34})\nonumber\\
c_{14}&=&r_{11}-r_{22}-r_{33}+r_{44}-2i\Re(r_{12})-2i\Im(r_{13})-2\Im(r_{14})-2\Im(r_{23})+2i\Im(r_{24})+2i\Re(r_{34})\nonumber\\
c_{41}&=&c_{14}^{*}\nonumber\\
c_{23}&=&r_{11}-r_{22}-r_{33}+r_{44}+2i\Re(r_{12})-2i\Im(r_{13})+2\Im(r_{14})+2\Im(r_{23})+2i\Im(r_{24})-2i\Re(r_{34})\nonumber\\
c_{32}&=&c_{23}^{*}\, ,\nonumber\\
%
%
d_{11}&=&1-2\Im(r_{14})+2\Im(r_{23})\nonumber\\
d_{22}&=&1+2\Im(r_{14})-2\Im(r_{23})\nonumber\\
d_{14}&=&r_{11}-r_{22}-r_{33}+r_{44}-2\Im(r_{14})-2\Im(r_{23})\nonumber\\
d_{23}&=&r_{11}-r_{22}-r_{33}+r_{44}+2\Im(r_{14})+2\Im(r_{23})\, .\nonumber
\eeqa

\end{itemize}

\section {Appendix 2}
We may also distinguish different routes for $C=A+B$, such that $A$, $B$ and $C$ commute with each other. As a simple example, let us take 
$C$ as the 2D unit operator 
\beq
\mathbf{1}= |+\ra\la +| + |-\ra\la -|\, ,
\eeq
where $\{|\pm\ra\}$ are eigenvectors of $\sigma_z$, 
and the operators $A=|+\ra\la +|$ and
$B=|-\ra\la -|$ have, each of them,  eigenvalues $+1$ or $0$.
If we assign preexisting values to each of the magnitudes that verify $C=A+B$ such that they just can take values $+1$ or $0$, it is impossible to assert if the magnitude $C$ was measured directly or, via $A+B$ as shown by the  table,
\beq
\begin{array}{ccc}
A & B & C \\
+1 & 0 & +1 \\
0 & +1 & +1\, .
\end{array}
\eeq
However, within the standard formalism of quantum mechanics, it is easy to discriminate both ways of measureing $C$. If our initial state is given by
\beq
\rho_i =\alpha|+\ra\la+| + \beta|+\ra\la-| + \gamma|-\ra\la+| + \delta|-\ra\la-|\, ,
\eeq
and we perform a L\"uders measurement of $C$, the state after the measurement does not change so
\beq
\rho_C =\rho_i =\alpha|+\ra\la+| + \beta|+\ra\la-| + \gamma|-\ra\la+| + \delta|-\ra\la-|\,.
\eeq
If we perform a L\"uders measurement of, first $A$ and then $B$, 
which is equivalent to making a von Neumann measurement of $C$  in the basis $\{|+\ra,|-\ra\}$,
we obtain
\beq
\rho_{A+B}=\alpha|+\ra\la+| + \delta|-\ra\la-|\, .
\eeq
Now, to distinguish the route followed, we just need the expectation value of, say $\sigma_x$, and compare the results,
\beqa
\la \sigma_x\ra_C &=& \beta +\gamma\nonumber\\
\la \sigma_x \ra_{A+B}&=&0\, .
\eeqa
%


\end{document}